\documentclass{article}
\usepackage{spconf,amsmath,graphicx,hyperref}
\usepackage{booktabs}
\usepackage{multirow} 
\usepackage{amssymb}
\usepackage{cite}
\usepackage{amsmath,amsfonts}
\usepackage{algorithmic}
\usepackage{algorithm}
\usepackage{array}
\usepackage[caption=false,font=normalsize,labelfont=sf,textfont=sf]{subfig}
\usepackage{tabularray}
\usepackage{textcomp}
\usepackage{stfloats}
\usepackage{url}
\usepackage{verbatim}
\usepackage{graphicx}
\usepackage{cite}
\usepackage{enumitem}
\usepackage{hyperref}
\usepackage{makecell}
\usepackage{etoolbox}
\robustify\bfseries
\usepackage{amssymb}
\usepackage{tabularx}
\usepackage{wasysym}
\usepackage{color,soul}
\usepackage{balance}
\usepackage{booktabs}
\usepackage{multirow}
\usepackage{siunitx}
\usepackage{arydshln} 
\usepackage{tabularray}
\usepackage{tblr-extras}

\usepackage{siunitx}
\usepackage{booktabs}
\usepackage{caption}    
\usepackage{flushend}
\usepackage[capitalise]{cleveref}

\makeatletter
\renewcommand\normalsize{%
   \@setfontsize\normalsize{9pt}{10.5pt}} 
\makeatother
\let\OLDthebibliography\thebibliography
\renewcommand\thebibliography[1]{
  \OLDthebibliography{#1}
  \setlength{\parskip}{0.5pt}
  \setlength{\itemsep}{1pt plus 0.3ex}
}
\title{PROSING: ZERO-SHOT FINE-GRAINED SINGING TECHNIQUE CONVERSION}
\name{Xinhao Guan$^1$, Jingyue Huang$^2$, Kangrui Chen$^1$, Julian McAuley$^2$, and Zhong-Qiu Wang$^1$}
\address{$^1$Southern University of Science and Technology, Shenzhen, China\\$^2$University of California San Diego, La Jolla, CA, USA}

\begin{document}
\ninept
\maketitle

\begin{abstract}
We study zero-shot fine-grained singing technique conversion (STC), which applies specified techniques to selected regions of an existing singing recording while preserving lyrics, melody, timing, and singer identity. 
Existing singing conversion and synthesis methods offer complementary strengths in source preservation and explicit technique control, but combining both in an audio-to-audio setting remains underexplored, especially with limited technique-annotated data. 
We propose \textit{ProSing}, a zero-shot fine-grained STC framework that adapts a pretrained flow-matching acoustic model with frame-level technique conditioning. 
To mitigate conditioning shortcuts during adaptation, we introduce a paired training strategy to learn generalizable technique representations. 
We further incorporate continuous pitch trajectories for explicit and expressive control of pitch-related techniques. 
Subjective and objective evaluations on unseen singers and source recordings show that ProSing achieves accurate technique control while preserving singing naturalness and singer similarity.
\end{abstract}


\section{Introduction}
\label{sec:intro}

Singing techniques shape vocal characteristics and contribute to vocal expressiveness~\cite{zhao2024sintechsvs}. 
Existing work related to technique control has mainly been explored through singing style conversion (SSC) and singing voice synthesis (SVS).
SSC modifies the style of an input recording while preserving singer identity, lyrics, and melody~\cite{violeta2026svcc,violeta2025serenade,wang2026s2voice}. Because these systems specify targets through reference-derived style representations rather than independently editable technique variables, modifying a target technique may also affect other style attributes such as rhythm and emotion~\cite{zhang2024tcsinger}.
%
In contrast, technique-controllable SVS provides explicit control over specific techniques~\cite{zhao2024sintechsvs,guo2025techsinger,zhang2025tcsinger,pan2025synthetic}, but synthesizes singing from lyrics and symbolic musical scores rather than directly modifying an existing recording~\cite{liu2022diffsinger}, making it difficult to preserve its original performance characteristics.
These limitations motivate audio-to-audio conversion with explicit, fine-grained control over singing techniques.

We formulate this problem as zero-shot fine-grained singing technique conversion (STC). Given a source singing recording not
deliberately performed with the requested target techniques, the goal is to apply specific techniques to selected regions while preserving the lyrics, source singer identity, timing, and underlying melody. Here, \textit{fine-grained} refers to temporally localized and multi-technique control, while \textit{zero-shot} refers to conversion for singers unseen during training. Following the six-technique inventory defined in GTSinger~\cite{zhang2024gtsinger}, we consider
four timbral techniques, including \textit{breathy voice}, \textit{falsetto}, \textit{mixed voice}, and
\textit{pharyngeal voice}, and two pitch techniques, \textit{vibrato} and
\textit{glissando}~\cite{zhao2024sintechsvs}.

Prior work has explored several forms of audio-to-audio singing-technique manipulation~\cite{luo2020techniquevae,oconnor2021zeroshot,chen2023fewshot,su2023stargan,choi2025vibesvc,hu2026boundary,hsu2026fabyol}, but two challenges remain. First, high-quality technique-annotated singing data remain scarce ~\cite{wilkins2018vocalset,zhang2024gtsinger}, as annotation requires expert knowledge and is costly to obtain
~\cite{zhang2024gtsinger}. Training an STC model from scratch on limited data may therefore pose challenges to acoustic quality and generalization~\cite{oconnor2021zeroshot}. Second, existing systems provide limited fine-grained controllability. Early approaches mainly perform global technique conversion~\cite{oconnor2021zeroshot,chen2023fewshot}, while recent methods with local control are restricted to a specific technique~\cite{choi2025vibesvc} or singers seen during training~\cite{hu2026boundary}. Zero-shot STC with local and multi-technique control remains underexplored.

To address the data challenge, we adapt the pretrained Vevo1.5 framework~\cite{zhang2025vevo,li2025amphion}, leveraging its large-scale speech and singing pretraining while retaining its tokenizer and fine-tuning the flow-matching acoustic model with frame-level technique conditions.
We retain its pretrained tokenizer for condition encoding and fine-tune its flow-matching acoustic model for STC.
However, naively incorporating technique conditions into the original training scheme can create conditioning shortcuts. The model may exploit technique cues from the timbre prompt and acoustic condition rather than learning technique representations that generalize across singers and recordings. We further find that discrete technique labels alone are insufficient to capture detailed pitch modulation in vibrato and glissando.

In this work, we propose \textbf{ProSing}, a zero-shot fine-grained STC framework. We adapt a large-scale pretrained voice imitation model and condition it on frame-level technique labels to enable temporally localized and multi-technique control. To mitigate conditioning shortcuts and encourage reliance on explicit technique conditions, ProSing uses paired training with randomized timbre prompting and phoneme-guided alignment. We further introduce continuous pitch trajectories to enable explicit modeling and control of pitch modulation. 

We evaluate ProSing on singers and source recordings unseen during training using subjective and objective metrics. ProSing substantially improves technique quality and controllability over a technique-controllable SVS baseline while maintaining high singing naturalness and singer similarity. Ablations further validate the proposed paired training and continuous pitch-trajectory conditioning. Our contributions are:
\begin{itemize}[leftmargin=*,nosep]
\item We formulate zero-shot fine-grained singing technique conversion with temporally localized and multi-technique control.
\item We develop a paired training strategy that enables pretrained acoustic models to learn explicit, generalizable technique control from limited annotated data.
\item We introduce continuous pitch-trajectory conditioning for expressive, fine-grained control of pitch-related techniques.
\item We provide subjective and objective evaluations of technique control and source preservation. Audio demos are available at \href{https://howardguan24.github.io/prosing-demo/}{https://howardguan24.github.io/prosing-demo/}.
\end{itemize}

\begin{figure*}[t]
  \centering
  \includegraphics[
    width=0.90\textwidth,
    height=0.35\textheight,
    keepaspectratio
]{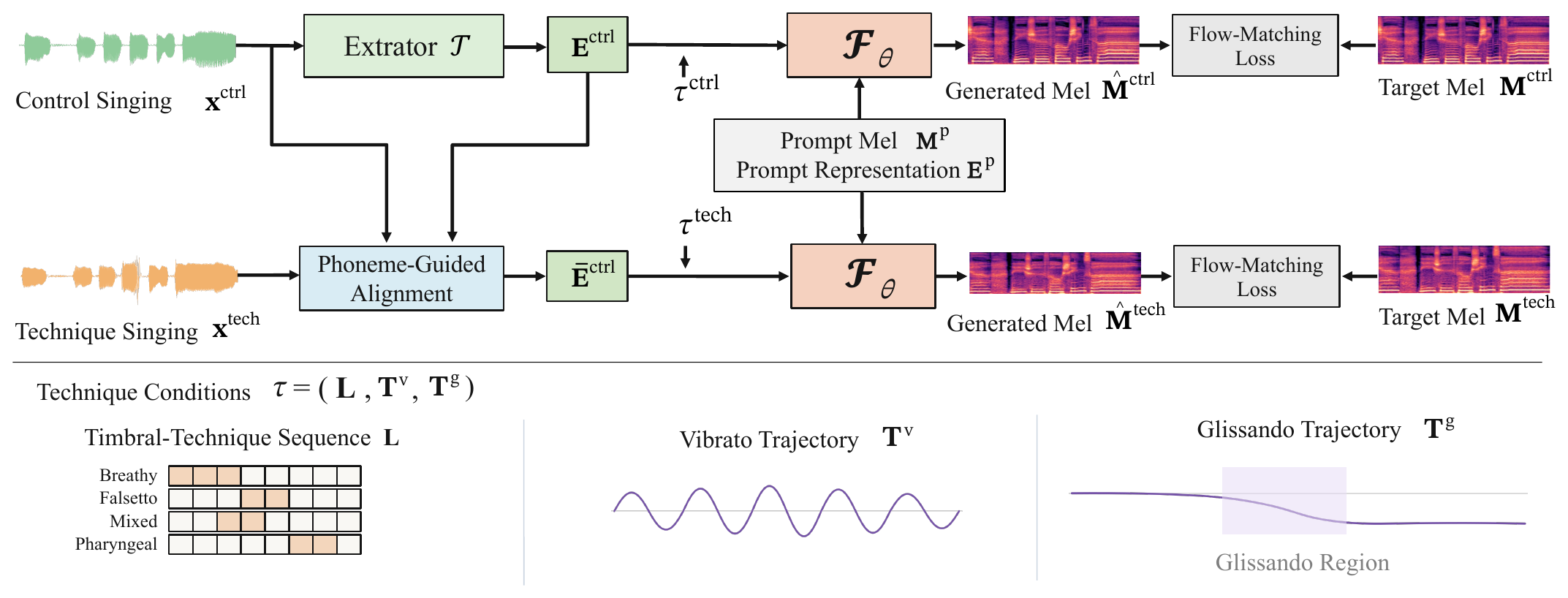}
  \vspace{-0.2cm}
  \caption{Overview of ProSing: paired training with phoneme-guided alignment (top) and fine-grained technique conditioning (bottom).}
  \label{fig:overview}
  \vspace{-0.3cm}
\end{figure*}

\section{Background}
\label{sec:related}

Vevo1.5 is a voice imitation framework released in Amphion~\cite{zhang2025vevo,li2025amphion,zhang2024amphion} that supports singing style conversion using a content-style tokenizer and a prompt-based flow-matching acoustic model\footnote{\url{https://github.com/open-mmlab/Amphion/tree/main/models/svc/vevosing}}.
The tokenizer encodes Whisper~\cite{radford2023whisper} and chromagram features~\cite{muller2011chroma} into $12.5$-Hz discrete tokens retaining linguistic content, melody, and performance-style information while reducing singer-timbre information~\cite{zhang2025vevo}.
The tokens are embedded and upsampled to $50$-Hz frame-level representations; we denote this extraction process by $\mathcal{T}$.


Given a timbre prompt $\mathbf{p}$ and source audio $\mathbf{a}$, let
$\mathbf{E}^{\mathrm{p}}=\mathcal{T}(\mathbf{p})$ and
$\mathbf{E}^{\mathrm{a}}=\mathcal{T}(\mathbf{a})$, with
$\widetilde{\mathbf{E}}=\mathbf{E}^{\mathrm{p}}\oplus\mathbf{E}^{\mathrm{a}}$
denoting temporal concatenation,
where $\oplus$ denotes temporal concatenation.
Let $\mathbf{M}^{\mathrm{p}}$ denote the prompt Mel spectrogram.
For the target Mel spectrogram $\mathbf{M}$, flow matching~\cite{lipman2023flowmatching} constructs an intermediate sample at time $t\in[0,1]$ as
\begin{equation}
\mathbf{M}^{t}=[1-(1-\sigma)t]\mathbf{Z}+t\mathbf{M},
\qquad
\mathbf{V}^{*}=\mathbf{M}-(1-\sigma)\mathbf{Z},
\end{equation}
where $\mathbf{Z}\sim\mathcal{N}(\mathbf{0},\mathbf{I})$ and $\sigma=10^{-5}$.
The noisy target $\mathbf{M}^{t}$ is concatenated with the prompt Mel as
$\widetilde{\mathbf{M}}^{t}=\mathbf{M}^{\mathrm{p}}\oplus\mathbf{M}^{t}$.
Finally, we combine the Mel and content-style conditions through learnable projections $\phi_M$ and $\phi_E$, and feed the resulting representation to the flow-matching Transformer (FMT) $\mathcal{F}_{\theta}$ to predict the flow velocity:
\begin{equation}
\mathbf{H}^{t} = 
\phi_M(\widetilde{\mathbf{M}}^{t})
+
\phi_E(\widetilde{\mathbf{E}}),
\qquad
\hat{\mathbf{V}}^{t} = 
\mathcal{F}_{\theta}(\mathbf{H}^{t},t).
\end{equation}
Training minimizes the masked $\ell_1$ loss between the predicted and target velocities $\hat{\mathbf{V}}^{t}$ and
$\mathbf{V}^{*}$ over valid target frames,
excluding prompt and padding regions.
During inference, the predicted velocity field is integrated from noise to generate the target Mel, which is then converted to waveform audio by a pretrained Vocos-based vocoder~\cite{siuzdak2024vocos}.
ProSing retains this generation pipeline while adapting its conditioning and training strategy for fine-grained STC.

\section{ProSing}
\label{sec:method}
Given an original singing recording $\mathbf{x}^{\mathrm{ori}}$, a same-singer timbre prompt $\mathbf{p}$, and technique condition $\tau$, ProSing generates
\begin{equation}
\hat{\mathbf{x}}
=
\operatorname{ProSing}\bigl(
\mathbf{x}^{\mathrm{ori}},
\mathbf{p};
\tau
\bigr),
\end{equation}
following
the specified techniques while preserving the source linguistic content, timing, singer identity, and underlying melody. 
$\mathbf{x}^{\mathrm{ori}}$ contains no deliberate performance of the requested techniques, while $\mathbf{p}$ comes from $\mathbf{x}^{\mathrm{ori}}$ or another recording by the same singer.

For training, we use paired adaptation with phoneme-guided alignment to reduce conditioning shortcuts. The technique condition $\tau=(\mathbf{L},\mathbf{T}^{\mathrm{v}},\mathbf{T}^{\mathrm{g}})$ combines frame-level timbral labels with continuous vibrato and glissando trajectories for local, multi-technique control. The overview of ProSing is shown in \cref{fig:overview}.



\subsection{Paired Training with Phoneme-Guided Alignment}
\label{ssec:paired}

Under the original Vevo1.5 training scheme~\cite{zhang2025vevo,li2025amphion,zhang2024amphion}, the FMT reconstructs the target Mel spectrogram conditioned on its content-style representation and a timbre prompt from the same target audio.
Since both conditions may already contain technique cues, directly adding explicit technique conditions can create conditioning shortcuts, allowing the model to reconstruct target techniques without fully relying on the explicit controls and limiting the generalization of learned technique representations.

We instead employ paired training using paired control and technique renditions.
Let $\mathbf{x}^{\mathrm{ctrl}}$ and $\mathbf{x}^{\mathrm{tech}}$ denote a
control rendition and its corresponding technique rendition, which share the
same singer, lyrics, and underlying melody.
Rather than conditioning each branch on its own content-style representation, $\mathbf{E}^{\mathrm{ctrl}}=\mathcal{T}(\mathbf{x}^{\mathrm{ctrl}})$ and $\mathbf{E}^{\mathrm{tech}}=\mathcal{T}(\mathbf{x}^{\mathrm{tech}})$, 
we derive the content-style conditions of both branches from $\mathbf{E}^{\mathrm{ctrl}}$.
The control branch uses $\mathbf{E}^{\mathrm{ctrl}}$ directly, while the technique branch uses its temporally aligned version defined below. 
This prevents technique cues from entering the technique branch through $\mathbf{E}^{\mathrm{tech}}$, encouraging reliance on explicit technique labels for reconstruction.

Although the paired renditions share the same linguistic content, their phoneme durations may differ. We therefore align $\mathbf{E}^{\mathrm{ctrl}}$ to the timing of the technique rendition $\mathbf{x}^{\mathrm{tech}}$.
We convert the annotated phoneme boundaries of both renditions to $50$-Hz frame positions and align their phoneme sequences using Levenshtein alignment~\cite{levenshtein1966}. Based on the matched phonemes, we construct a piecewise-linear mapping from the technique timeline to the control timeline.
For each technique frame $i$, let $\gamma_i$ denote its corresponding,
possibly fractional, position on the control timeline.
For $\gamma_i=k+\alpha$, where $k=\lfloor\gamma_i\rfloor$ and
$0\leq\alpha<1$, the aligned control representation is obtained by linear interpolation:
\begin{equation}
\bar{\mathbf{E}}^{\mathrm{ctrl}}_i
=
(1-\alpha)\mathbf{E}^{\mathrm{ctrl}}_k
+
\alpha\mathbf{E}^{\mathrm{ctrl}}_{k+1}.
\end{equation}
Thus, $\bar{\mathbf{E}}^{\mathrm{ctrl}}$ follows the timing of the technique rendition while retaining control-derived content-style information, providing phoneme-aligned conditioning.

To further reduce reliance on utterance-specific acoustic cues in the timbre prompt, both branches share a prompt randomly cropped from another control recording by the same singer in each epoch.
We also construct composite-technique data to expose the model to within-utterance transitions between timbral-technique states. Each example concatenates two same-singer segments from different technique groups, sampled from either the technique or control rendition.

\subsection{Fine-Grained Technique Conditioning}
\label{ssec:conditioning}

The technique condition
$\tau=(\mathbf{L},\mathbf{T}^{\mathrm{v}},\mathbf{T}^{\mathrm{g}})$
combines categorical control for timbral techniques with continuous control for pitch techniques.
For an input with $N$ acoustic frames,  $\mathbf{L}\in\{0,1\}^{N\times 4}$ specifies the frame-level activation of breathy voice, falsetto, mixed voice, and pharyngeal voice, allowing multiple techniques to overlap.
Vibrato and glissando are controlled by continuous trajectories
$\mathbf{T}^{\mathrm{v}},\mathbf{T}^{\mathrm{g}}\in\mathbb{R}^{N}$.
Unlike binary labels, which indicate only technique activity, these trajectories
explicitly specify frame-level pitch modulation, such as vibrato phase and
glissando direction, avoiding averaging over different pitch realizations.


We extract $F_0$ with pYIN~\cite{mauch2014pyin} and express it in cents as
$\mathbf{f}$.
The vibrato trajectory removes slow pitch variation with a centered $260$-ms moving average and retains the vibrato-frequency component:
\begin{equation}
\mathbf{T}^{\mathrm{v}}
=
\mathcal{B}\big(
\mathbf{f}-\mathcal{A}(\mathbf{f})
\big),
\end{equation}
where $\mathcal{A}$ is the moving-average operator and $\mathcal{B}$ a
$3$--$9$ Hz band-pass filter spanning typical trained-singer vibrato rates ($5$--$7$ Hz)~\cite{nusseck2007effects}.
Unvoiced frames are zeroed, and $F_0$ sequences with excessive octave-scale
tracking errors or insufficient voiced frames are discarded.

For glissando, let $\mathbf{f}^{\mathrm{tech}}$ denote the vibrato-removed
pitch trajectory of the technique rendition and
$\bar{\mathbf{f}}^{\mathrm{ctrl}}$ the 
corresponding vibrato-removed control
trajectory warped to the technique timeline using the phoneme-guided alignment
in Sec.~\ref{ssec:paired}.
We define
\begin{equation}
\mathbf{T}^{\mathrm{g}}
=
\mathbf{f}^{\mathrm{tech}}
-
\bar{\mathbf{f}}^{\mathrm{ctrl}}.
\end{equation}
Frames with invalid $F_0$ estimates or pitch differences exceeding one octave are
set to zero, and unreliable trajectory pairs are discarded.

In ProSing, the content-style condition is
\begin{equation}
\widetilde{\mathbf{E}}
=
\begin{cases}
\mathbf{E}^{\mathrm{p}}\oplus\mathbf{E}^{\mathrm{ctrl}},
& \text{control reconstruction},\\
\mathbf{E}^{\mathrm{p}}\oplus\bar{\mathbf{E}}^{\mathrm{ctrl}},
& \text{technique reconstruction},\\
\mathbf{E}^{\mathrm{p}}\oplus\mathbf{E}^{\mathrm{ori}},
& \text{inference},
\end{cases}
\end{equation}
where $\mathbf{E}^{\mathrm{ori}}=\mathcal{T}(\mathbf{x}^{\mathrm{ori}})$.
The technique conditions are zero-padded over the prompt frames, yielding
$\widetilde{\mathbf{L}}$, $\widetilde{\mathbf{T}}^{\mathrm{v}}$, and
$\widetilde{\mathbf{T}}^{\mathrm{g}}$.
The Vevo1.5 FMT input is then augmented as
\begin{equation}
\mathbf{H}^{t}
=
\phi_M(\widetilde{\mathbf{M}}^{t})
+
\phi_E(\widetilde{\mathbf{E}})
+
\phi_L(\widetilde{\mathbf{L}})
+
\phi_{\mathrm{v}}(\widetilde{\mathbf{T}}^{\mathrm{v}})
+
\phi_{\mathrm{g}}(\widetilde{\mathbf{T}}^{\mathrm{g}}),
\end{equation}
where $\phi_L$, $\phi_v$, and $\phi_g$ are learnable projections.
Let $\mathbf{H}_{\mathrm{ctrl}}^{t}$ and
$\mathbf{H}_{\mathrm{tech}}^{t}$ denote the fused inputs for two branches, and $\mathbf{V}_{\mathrm{ctrl}}^{*}$ and
$\mathbf{V}_{\mathrm{tech}}^{*}$ their target velocities defined as in \cref{sec:related}. With $\operatorname{Mask}(\cdot)$ excluding prompt and
padding regions, the paired objective is
\begin{align}
\mathcal{L}_{\mathrm{pair}}
={}&
\left\|
\operatorname{Mask}\left(
\mathcal{F}_{\theta}(\mathbf{H}_{\mathrm{ctrl}}^{t},t)
-
\mathbf{V}_{\mathrm{ctrl}}^{*}
\right)
\right\|_1
\nonumber\\
&+
\left\|
\operatorname{Mask}\left(
\mathcal{F}_{\theta}(\mathbf{H}_{\mathrm{tech}}^{t},t)
-
\mathbf{V}_{\mathrm{tech}}^{*}
\right)
\right\|_1 .
\end{align}

\section{Experimental Setup}

\subsection{Datasets}

We use the technique-annotated GTSinger dataset~\cite{zhang2024gtsinger} for training, which provides paired control and technique renditions and contains $80.59$ hours of singing from $20$ professional singers across $9$ languages. To preserve zero-shot singer generalization of Vevo1.5 after fine-tuning, we additionally include $53.06$ hours of singing from $58$ singers in OpenSinger~\cite{huang2021multi} and M4Singer~\cite{zhang2022m4singer}, with $38$ singers held out for evaluation. Since these samples lack technique annotations, we condition them on their original content-style representations and same-singer timbre prompts, retaining only naturally extracted vibrato trajectories. The GTSinger and additional-data streams are balanced at a $1:1$ ratio based on the total number of $50$-Hz frames.

\subsection{System Configurations}

ProSing is initialized from the official Vevo1.5 checkpoint~\cite{zhang2025vevo,li2025amphion,zhang2024amphion}, pretrained on approximately $101$k hours of speech and $7$k hours of singing.
We freeze the Whisper and chromagram extractors, content-style tokenizer, and vocoder, while fine-tuning the FMT together with the newly introduced technique-conditioning modules.

We train on $4$ NVIDIA RTX A6000 GPUs using dynamic batches of up to $5{,}000$ $50$-Hz frames and $50$ utterances per GPU. We use the AdamW optimizer~\cite{loshchilov2019decoupled} with a peak learning rate of $10^{-5}$, $2{,}000$ warm-up steps, inverse-square-root decay, and bfloat16 precision.
The first five ablation variants in~\cref{tab:ablation} are trained for $400$k steps, while the longer-training variant uses $517.2$k steps, all on a single GPU. The final ProSing model with additional singing data is trained on $4$ GPUs with approximately the same total training-frame exposure of the longer-training checkpoint.

Inference uses $32$-step Euler integration with the 
CFG coefficient set to $1.0$, following the Vevo1.5 implementation, and the pretrained Vevo1.5 vocoder for waveform synthesis.

\subsection{Baselines}

We compare ProSing with TCSinger~\cite{zhang2024tcsinger} and Vevo1.5~\cite{zhang2025vevo,li2025amphion}, restricting baselines to systems with compatible technique-control settings and publicly available checkpoints for consistent quantitative comparison. We use the official STARS model~\cite{guo2025stars} to obtain both the symbolic annotations required for TCSinger inference and the technique predictions used for objective evaluation.

TCSinger~\cite{zhang2024tcsinger} is a zero-shot technique-controllable SVS system; we use its official bilingual checkpoint trained on approximately $294$ hours of data.
To adapt TCSinger to our conversion setting, we first extract score conditions from the source recording using STARS~\cite{guo2025stars}, then provide them together with the same $3$-second source-singer timbre prompt and target technique labels used by ProSing.

For Vevo1.5~\cite{zhang2025vevo,li2025amphion}, we use the same source recording and timbre prompt, while providing a GTSinger recording exhibiting the target technique as the style reference.
Since Vevo1.5 performs global style conversion rather than localized technique conversion, we evaluate it only on single-technique cases.

\begin{table*}[t]
\centering
\caption{Subjective and objective evaluation of ProSing and the baselines on general conversion, technique control, and source preservation. MOS scores are reported as case-averaged means with $95$\% confidence intervals, shown as $\pm$ half-widths. CER and $F_0$-RMSE are evaluated on $36$ Chinese single-technique cases and $24$ timbral-technique cases, respectively. The source recordings have a CER of $23.5\%$.}
\label{tab:cross_system}
\vspace{-0.1cm}
\footnotesize
\setlength{\tabcolsep}{5pt}
\begin{tabular}{lcccccccc}
\toprule
& \multicolumn{2}{c}{\textbf{Overall Quality}} & \multicolumn{2}{c}{\textbf{Single-Technique}} & \multicolumn{2}{c}{\textbf{Multi-Technique}} & \multicolumn{2}{c}{\textbf{Source Preservation}} \\
\cmidrule(lr){2-3} \cmidrule(lr){4-5} \cmidrule(lr){6-7} \cmidrule(lr){8-9}
\textbf{Method} & N-MOS $\uparrow$ & S-MOS $\uparrow$ & TQ-MOS $\uparrow$ & TA-MOS $\uparrow$ & TQ-MOS $\uparrow$ & TA-MOS $\uparrow$ & CER (\%) $\downarrow$ & $F_0$-RMSE (cents) $\downarrow$ \\
\midrule
TCSinger~\cite{zhang2024tcsinger} & $1.48 \pm 0.14$ & $1.64 \pm 0.15$ & $1.41 \pm 0.12$ & $1.60 \pm 0.15$ & $1.41 \pm 0.19$ & $1.61 \pm 0.25$ & $64.1$ & $413.4$ \\
Vevo1.5~\cite{zhang2025vevo,li2025amphion} & $2.41 \pm 0.28$ & $2.68 \pm 0.25$ & $2.24 \pm 0.25$ & $2.26 \pm 0.25$ & -- & -- & $39.5$ & $332.8$ \\
ProSing (ours) & $\mathbf{3.36} \pm 0.22$ & $\mathbf{3.13} \pm 0.22$ & $\mathbf{3.06} \pm 0.22$ & $\mathbf{3.27} \pm 0.24$ & $\mathbf{2.91} \pm 0.53$ & $\mathbf{3.31} \pm 0.51$ & $\mathbf{27.6}$ & $\mathbf{166.7}$ \\
\bottomrule
\end{tabular}
\end{table*}

\begin{table*}[t]
\centering
\caption{Objective evaluation of ProSing ablation variants for different training strategies. Each ``+'' row incrementally adds the indicated component to the preceding configuration.}
\vspace{-0.1cm}
\label{tab:ablation}
\footnotesize
\setlength{\tabcolsep}{6pt}
\begin{tabular}{lcccccc}
\toprule
& \multicolumn{3}{c}{\textbf{Technique Following ($F_1 \uparrow$)}} & \multicolumn{1}{c}{\textbf{Trajectory Following}} & \multicolumn{2}{c}{\textbf{General Performance}} \\
\cmidrule(lr){2-4} \cmidrule(lr){5-5} \cmidrule(lr){6-7}
\textbf{Training Variant} & Single-Technique & Multi-Technique & Pitch-Technique & $\Delta F_0$-CCC $\uparrow$ & SingMOS-Pro $\uparrow$ & SpkSim $\uparrow$ \\
\midrule
Label-only fine-tuning & $0.143$ & $0.117$ & $0.078$ & -- & $4.441$ & $\mathbf{0.753}$ \\
\quad + Paired training & $0.265$ & $0.194$ & $0.102$ & -- & $4.424$ & $0.538$ \\
\quad + Phoneme-guided alignment & $0.389$ & $0.228$ & $0.094$ & -- & $\mathbf{4.448}$ & $0.533$ \\
\quad + Composite-technique data & $0.373$ & $0.221$ & $0.097$ & -- & $4.411$ & $0.528$ \\
\quad + Continuous trajectories & $0.410$ & $0.269$ & $0.162$ & $\mathbf{0.756}$ & $4.342$ & $0.526$ \\
\quad + Longer training & $\mathbf{0.439}$ & $\mathbf{0.291}$ & $\mathbf{0.195}$ & $0.739$ & $4.329$ & $0.521$ \\
\quad + Additional singing data & $0.374$ & $0.216$ & $0.184$ & $0.742$ & $4.367$ & $0.631$ \\
\bottomrule
\end{tabular}
\vspace{-0.3cm}
\end{table*}

\subsection{Evaluation Protocol and Metrics}

\noindent\textbf{Cross-system Comparison.}
In the subjective listening test, listeners independently rate four perceptual aspects of each technique-converted sample on a $5$-point Likert scale:
\begin{itemize}[leftmargin=*, itemsep=0pt, topsep=1pt]
\item Naturalness MOS (N-MOS): overall naturalness.
\item Singer Similarity MOS (S-MOS): source-singer similarity.
\item Technique Quality MOS (TQ-MOS): naturalness and quality of the requested techniques.
\item Technique Accuracy MOS (TA-MOS): accuracy of the requested techniques in the specified regions.
\end{itemize}
Evaluation cases are drawn from various singers unseen during training and
include both single- and multi-technique conversions with balanced technique
coverage.
A total of $28$ participants with diverse singing backgrounds provide the ratings. For each
single technique, each system receives at least $36$ ratings per metric, while the
multi-technique evaluation includes $84$ ratings per system and metric.
Scores are averaged across cases and reported with $95\%$ confidence intervals.
For objective evaluation, we assess source preservation using Whisper-based corpus-level
CER~\cite{radford2023whisper} for linguistic content and $F_0$-RMSE in cents
~\cite{mauch2014pyin,davis1980mfcc,sakoe1978dtw} for pitch preservation.
\\

\noindent\textbf{Ablation Study.}
We use three groups of objective metrics. For technique following, we report phoneme-level $F_1$ scores from STARS technique
predictions~\cite{guo2025stars} for single-technique, multi-technique, and
pitch-technique control. For reference, STARS achieves $0.726$ average $F_1$ across the six techniques on real GTSinger recordings, with only $0.380/0.534$ on vibrato/glissando.
For trajectory following, $\Delta F_0$-CCC~\cite{lin1989ccc} measures the agreement between the specified continuous pitch trajectory and the $F_0$ difference between matched outputs generated with and without trajectory conditioning.
For general performance, we report SingMOS-Pro~\cite{tang2026singmospro} for
singing quality and RawNet3 cosine similarity~\cite{jung2022rawnet3} (SpkSim)
for singer similarity.

For timbral-technique evaluation, we render $30$ held-out source recordings with low
STARS-predicted timbral-technique density under each of the $4$ technique
commands.
Multi-technique evaluation uses a subset of these sources with time-varying
combinations of two distinct technique states.
Pitch-technique following is evaluated separately on vibrato and glissando
cases.

\section{Evaluation Results}

\noindent\textbf{Cross-system Comparison.}
\cref{tab:cross_system} shows that ProSing outperforms both baselines across all subjective and objective metrics, achieving the highest mean subjective scores and the lowest CER and $F_0$-RMSE. Its strong naturalness and singer similarity indicate that improved technique controllability does not compromise perceptual quality or source identity. ProSing also maintains strong performance in the more challenging multi-technique setting, demonstrating effective localized control with strong source preservation.
\\

\noindent\textbf{Ablation Study.}
\cref{tab:ablation} shows that label-only fine-tuning yields weak Single-/Multi-Technique $F_1$ of $0.143/0.117$. Paired training improves these scores to $0.265/0.194$, and phoneme-guided alignment further raises them to $0.389/0.228$, supporting reduced conditioning shortcuts and stronger explicit technique following. Composite-technique data increases training diversity but brings no immediate STARS-based $F_1$ gain.

Continuous trajectories improve Pitch-Technique $F_1$ from $0.097$ to $0.162$ and achieve a $\Delta F_0$-CCC of $0.756$, supporting continuous pitch control. Longer training further raises Single-/Multi-/Pitch-Technique $F_1$ to $0.439/0.291/0.195$. Additional singing data improves SpkSim from $0.521$ to $0.631$, with some trade-off in technique following. Overall, the ablations reveal complementary effects on technique following, pitch control, and singer similarity.

\section{Conclusions}

We propose ProSing, a zero-shot fine-grained STC framework that adapts a
large-scale pretrained flow-matching acoustic model with limited
technique-annotated singing data for singing technique control. ProSing employs frame-level technique
conditioning for fine-grained control and a paired training strategy to reduce conditioning shortcuts and improve technique following. 
Continuous pitch trajectories are further incorporated for explicit control of pitch-related techniques. 
Subjective and objective evaluations demonstrate strong technique controllability and source preservation, with ablation studies validating the effectiveness of the proposed adaptation strategies.
Future work will extend ProSing to support direct conversion from one singing technique to another.

\clearpage
{\footnotesize
\bibliographystyle{IEEEtran}
\bibliography{refs.bib}
}

\end{document}